\documentclass[final,3p,times]{elsarticle}
\usepackage{amssymb}
\usepackage{amsmath,bm,graphicx}

\journal{Journal of Computational Physics}

\begin{document}

\begin{frontmatter}



\title{High dimensional linear inverse modelling}


\author{Fenwick C. Cooper}
\ead{Fenwick@LittleStick.com}

\address{Atmospheric, Oceanic and Planetary Physics, Department of Physics, University of Oxford, Oxford, OX1 3PU, UK}

\cortext[cora]{Corresponding author. Tel.: +44 7531 290 687}

\begin{abstract}

We introduce and demonstrate two linear inverse modelling methods for systems of stochastic ODE's with accuracy that is independent of the dimensionality (number of elements) of the state vector representing the system in question. Truncation of the state space is not required. Instead we rely on the principle that perturbations decay with distance or the fact that for many systems, the state of each data point is only determined at an instant by itself and its neighbours. We further show that all necessary calculations, as well as numerical integration of the resulting linear stochastic system, require computational time and memory proportional to the dimensionality of the state vector. % for each integration step.
%One of these calculations is a method of generating random numbers with a specified covariance, or spatial relationship.
%, $O(nd)$ in big O notation. 
%We demonstrate our method by application to a single pressure level (857.1 mb) of the ERA-interim atmospheric temperature reanalysis data set. The state vector of our data subset has a dimensionality of $d=41346$ points taken over a time period of 21 years represented by $n=30681$ points in time.

\end{abstract}

\begin{keyword}

Inverse modelling \sep Regression \sep Fluctuation-dissipation \sep Stochastic \sep Correlated random numbers



\end{keyword}

\end{frontmatter}


%\section{}
%\label{}

%%%%%%%%%
% New Section %
%%%%%%%%%

\section{Introduction}

Consider the linear stochastic system of ordinary differential equations
\begin{equation}
%\text{d}\mathbf{x} = \mathbf{Bx}\text{d}t + \mathbf{Q} \text{d}\mathbf{w}
\frac{\text{d}\mathbf{x}}{\text{d}t}=\mathbf{Bx}+\bm{\xi}.
\label{linearModel:eqn}
\end{equation}
where $\mathbf{x}$ is a $d$ dimensional vector that contains the state of the system at a particular time $t$, $\mathbf{B}$ is a matrix of constant coefficients and $\bm{\xi}$ denotes a vector white noise process. The notation is chosen following \cite{Penland95}. We consider the case were the eigenvalues of $\mathbf{B}$ all have negative real parts and
\begin{equation}
\mathbf{Q}=\left< \bm{\xi \xi}^{\text{T}} \right>
\end{equation}
is the noise covariance matrix which is symmetric positive definite. The angled brackets $\left< \dots \right>$ denote the expectation value and $\text{T}$ denotes the matrix transpose. Given a time series of $n$ data points $\mathbf{X}_i$, $i=1 \dots n$, each representing $\mathbf{x}$ at a particular time $t$, our aim is to find $\mathbf{B}$ and $\mathbf{Q}$.

%%%%%%%%
% Subsection %
%%%%%%%%

\subsection{Applications}

The linear system (\ref{linearModel:eqn}) with $\mathbf{B}$ and $\mathbf{Q}$ found from data has been applied for many years to approximate the dynamics of non-linear systems \cite{Penland89}. In particular, analysis of the surface temperatures in the pacific (e.g. \cite{Penland95}, \cite{Newman07}) and atlantic (e.g. \cite{Hawkins09},\cite{Zanna12}) oceans have been studied, with extensions to the sub surface dynamics (e.g. \cite{Newman11}). A closely related approach is to solve the system of ocean governing equations on a computationally feasible grid, necessitating a higher viscosity. One or both of the terms in (\ref{linearModel:eqn}) are added to the right hand side of the governing equations to approximate the sub-grid scale flow (e.g. \cite{Hasselmann76}, \cite{Frederiksen97}, \cite{Achatz99}, \cite{Berloff05}, \cite{Frederiksen06}, \cite{Grooms13}, \cite{Kitsios13}, \cite{Cooper15b}). The author's motivation for this work is to find an improved estimate of $\mathbf{B}$ and $\mathbf{Q}$ and use (\ref{linearModel:eqn}) for this purpose.

The time averaged statistics of a fluid flow may be found by integrating the governing equations for a sufficient length of time. Rather than solving an equation governing the instantaneous flow, several authors have considered solving the equations for the statistics (e.g. \cite{Farrell03},\cite{Marston08}, \cite{Srinivasan12}, \cite{Tobias13}, \cite{Bakas14}). Such attempts require neglecting all of the cumulants beyond the first two, or parameterising missing terms with a linear stochastic term. The only system with two cumulants is a Gaussian system \cite{Lukacs70} and (\ref{linearModel:eqn}) is a system that is capable of replicating a Gaussian probability density function (PDF). If such statistical equations are applicable then a method of closure is to use the statistics measured from a flow to estimate $\mathbf{B}$ and $\mathbf{Q}$, and hence the governing system (\ref{linearModel:eqn}).

If (\ref{linearModel:eqn}) with appropriate parameters can be used as an accurate model of the Earth's oceans or atmosphere (or sub-grid model), then it may be applied to improve estimates of climate change (e.g. \cite{Gritsun07}), although the utility of such linear estimates may be qualitative \citep{Cooper15a} . Adding $\delta \mathbf{f}$ to the right hand side of (\ref{linearModel:eqn}) to represent a constant forcing causes the time mean, or climatological mean, of $\mathbf{x}$ to change. Denoting this change $\left< \delta \mathbf{x} \right>$ we get
\[
\left< \delta \mathbf{x} \right> = -\mathbf{B}^{-1} \delta \mathbf{f}.
\]
The form of forcing given a particular response can also be found by rearranging to get
\[
\delta \mathbf{f} = -\mathbf{B} \left< \delta \mathbf{x} \right>.
\]

Accurate representation of (\ref{linearModel:eqn}) also has potential for use with statistical significance testing. A common hypothesis to test is that some data is significantly different from uncorrelated Gaussian random noise. The appropriate test in this case is Students t-test. The assumption of independence required by Students t-test is not satisfied in the case of data that is correlated in time and the test is not appropriate. There are alternatives (see \cite{vonStorch99}), one is to use Monte-Carlo integration to test that the time series of data is significantly different from a first order auto-regressive process, the discrete analogue of (\ref{linearModel:eqn}) with dimension $d=1$. However much of the science of the ocean and atmosphere involves the analysis of large data sets that are both spatially and temporarily correlated. Both spatial and temporal correlation can be included in the null-hypothesis model by considering (\ref{linearModel:eqn}) with $d>1$.

%The key to fast numerical integration of (\ref{linearModel:eqn}) is a fast method of generating a vector of random numbers with a specified covariance. Current methods include for example [[Cite fourier method]] used in XXX and uniform choice of a decorrelation length scale (e.g. \cite{Buizza99}, \cite{Palmer09}) used in weather forecast models. These methods do not currently allow arbitrary control over the spatial correlation of the noise process and represent examples where a fast method could be useful. [[XXX Review the literature a bit better]]

%%%%%%%%
% Subsection %
%%%%%%%%

\subsection{Linear Inverse Modelling (LIM)}

Multiplying (\ref{linearModel:eqn}) by $\mathbf{x}(0)^\text{T}$, taking the expectation value and solving the system of ordinary differential equations, gives the lag $\tau$ covariance matrix
\begin{align}
\mathbf{C}(\tau)
&=\exp\left( \mathbf{B} \tau \right) \mathbf{C}(0) \label{covExp:eqn} \\
&=\left< \mathbf{x}(\tau) \mathbf{x}(0)^{\text{T}} \right> \\
& \approx \frac{1}{n-m} \sum_{i=1}^{n-m} \mathbf{X}_{i+m} \mathbf{X}_{i}^{\text{T}},
\label{CtauFromData:eqn}
\end{align}
where the matrix exponential,
\[
\exp \left( \mathbf{B} \right) \equiv \sum_{k=0}^{\infty} \frac{1}{k!} \mathbf{B}^k,
\]
is used and $m$ is the number of data points within the time $\tau$. The covariance matrices may be estimated from the data and by rearranging (\ref{covExp:eqn}), $\mathbf{B}$ can be expressed in terms of the covariance matrices,
\begin{equation}
\mathbf{B}=\frac{1}{\tau}\log\left[ \mathbf{C}(\tau) \mathbf{C}(0)^{-1} \right]
\label{linearFDT:eqn}
\end{equation}
where the matrix logarithm (principal value of the inverse matrix exponential) is used and $\mathbf{B}$ is independent of $\tau$. This expression is sometimes referred to as the linear fluctuation dissipation relation or the linear fluctuation dissipation theorem. Integrating (\ref{covExp:eqn}) with respect to $\tau$ between zero and infinity, we get
\begin{equation}
\mathbf{B}=-\left( \int_0^{\infty} \mathbf{C}(\tau) \mathbf{C}(0)^{-1} d\tau \right)^{-1}
\label{gaussianFDT:eqn}
\end{equation}
which is sometimes referred to as the Gaussian fluctuation dissipation theorem (FDT) because it may be derived by assuming that the system in question has a Gaussian probability density function (PDF) instead of assuming the linear stochastic system (\ref{linearModel:eqn}) (e.g. \cite{Risken84}, \cite{Majda05}, \cite{Cooper11}).
%
% B=-(Ci.C0^-1)^-1
% B=-C0.Ci^-1
% B.Ci=-C0
% Ci^T . B^T = -C0
%
Having found $\mathbf{B}$, the noise covariance matrix $\mathbf{Q}$ may be found using the Lyapunov equation (e.g. \cite{Penland89})
\begin{equation}
\mathbf{BC}(0)+\mathbf{C}(0)\mathbf{B}^{\text{T}}+\mathbf{Q}=0.
\label{Lyapunov:eqn}
\end{equation}

%%%%%%%%
% Subsection %
%%%%%%%%

\subsection{Truncation of the data set}

In (\ref{linearFDT:eqn}) and (\ref{gaussianFDT:eqn}) the inverse of the covariance matrix $\mathbf{C}(0)^{-1}$ appears.
%In order to find it's inverse, the covariance matrix cannot be singular. Given that we are estimating $\mathbf{C}(0)$ using the data points $X$, and it will therefore not be exact, what are our requirements upon the data set?
%
For a non-singular $\mathbf{C}(0)$ (so that its inverse can be found) it is required that the number of measurements in time, $n$, is greater than the dimensionality, $d$, of the data set, $n>d$. Thus, and somewhat paradoxically, the more data that is collected at each interval of time (large $d$), the longer the data must be collected for (large $n$). This is a fundamental problem with (\ref{linearFDT:eqn}) and (\ref{gaussianFDT:eqn}) and means that the length of time data is required to be recorded over, can be impossibly long. A less serious problem is that if two data points behave in a similar way, because they reflect two measurements of a similar physical quantity, then two of the rows of $\mathbf{C}(0)$ are also similar and it is close to singular. It may take a lot of data to accurately characterise the difference between the two points and achieve a numerically non-singular $\mathbf{C}(0)$.

Another issue is the computational cost. Computation of the matrix inverse in (\ref{linearFDT:eqn}) and (\ref{gaussianFDT:eqn}), the matrix logarithm in (\ref{linearFDT:eqn}) and the general multiplication of $\mathbf{B}$ and $\mathbf{C}(0)$ in (\ref{Lyapunov:eqn}) each take of the order of $d^3$ floating point operations on a computer (e.g. \cite{Press07},\cite{Cheng01}). Having found $\mathbf{B}$ and $\mathbf{Q}$, there is also the problem of numerical integration of the linear system (\ref{linearModel:eqn}). For the evaluation of $\bm{\xi}$, the eigenvalues and vectors of $\mathbf{Q}$ must be found once, taking of the order of $d^3$ operations, and in general both the $\mathbf{Bx}$ and $\bm{\xi}$ terms take $d^2$ operations per time step, see \ref{RandomTerm:sect}. Thus, for large $d$ the computational cost of either finding $\mathbf{B}$ and $\mathbf{Q}$ or integrating (\ref{linearModel:eqn}) becomes prohibitive.

The solution to the problem of $d$ being too large, is to truncate the data set to a lower dimensionality $d'$, where $d'$ is sufficiently small for practical use. This is typically achieved by finding the leading $d'$ eigenvalues and eigenvectors of $\mathbf{C}(0)$ and setting the remaining $d-d'$ eigenvalues to zero. The data is then transformed into the space defined by the matrix of the leading $d'$ eigenvectors, $\mathbf{V}_{\text{s}}$, commonly referred to as Empirical Orthogonal Function (EOF) space (e.g. \cite{Jolliffe02}). Giving
\begin{equation}
\mathbf{C}_{\text{s}}(t)=\mathbf{V}_{\text{s}}^{\text{T}} \mathbf{C}(t) \mathbf{V}_{\text{s}}, \qquad
\mathbf{B}_{\text{s}}=\mathbf{V}_{\text{s}}^{\text{T}} \mathbf{B} \mathbf{V}_{\text{s}} \qquad
\text{and} \qquad
\mathbf{Q}_{\text{s}}=\mathbf{V}_{\text{s}}^{\text{T}} \mathbf{Q} \mathbf{V}_{\text{s}}.
\label{EOFTruncation:eqn}
\end{equation}
Here the subscript s denotes the truncated matrices. The problem of inverting the $d' \times d'$ matrix $\mathbf{C}_{\text{s}}(0)$ then requires that $n>d'$, calculation of $\mathbf{B}_{\text{s}}$ and $\mathbf{Q}_{\text{s}}$ takes of the order of $d'^3$ operations and integration of the truncated version of (\ref{linearModel:eqn}) takes of the order of $d'^2$ operations per time step. An estimate of the full $\mathbf{B}$ and $\mathbf{Q}$ matrices may then be made by performing the inverse of (\ref{EOFTruncation:eqn}). The assumption is that the most important processes have the largest variance. Even if this is true, neglecting the least variable processes combined with inaccuracies in the estimation of $\mathbf{V}_{\text{s}}$ introduces bias into the estimation of $\mathbf{B}$ and $\mathbf{Q}$. For a chaotic system where each point is governed by the same rules as its neighbours, truncation in EOF space may not be the most appropriate truncation to make. In this case a localised truncation can be optimal \cite{Cooper13}.

%%%%%%%%
% Subsection %
%%%%%%%%

\subsection{Paper Outline}

The purpose of this paper is to introduce two alternative methods that do not require truncation in EOF space, instead relying on the assumption of locality and using the fact that for many problems, $\mathbf{B}$ is sparse. Locality is defined by assuming that elements of $\mathbf{C}(\tau)$ relating two points with a distance greater than some critical value, may be set to zero. This is the same as assuming that if there are no significant correlations at a lag $\tau$ between two variables, there is no evidence that they have any significant relation at this lag, and are therefore assumed to be effectively independent. With these assumptions, the accuracy of any estimate of $\mathbf{B}$ becomes independent of the dimensionality $d$ of the state vector $\mathbf{x}$. Practical results of this approach are that bias and smoothing due to EOF truncation are eliminated and that less data is required for a given accuracy.

In section \ref{method:sect} we describe two methods of finding a local $\mathbf{B}$ and $\mathbf{Q}$ that have an accuracy independent of $d$.
%In \ref{RandomTerm:sect} a method is presented for the fast calculation of correlated noise, which enables the evaluation of  (\ref{linearModel:eqn}) with of the order $d$ computations per time step.
In section \ref{testModel:sect} we introduce a test model to evaluate the ability of  the algorithms presented in section \ref{method:sect}. In section \ref{results:sect}, using limited data (often $n\ll d$) from the test models, with $d$ between 2 and $2^{16}$, (65536), the performance of the algorithms is demonstrated. Our conclusions are described in section \ref{conclusions:sect}.

%%%%%%%%%
% New Section %
%%%%%%%%%

\section{The method}
\label{method:sect}

\subsection{The local Gaussian FDT}
\label{localGaussianFDT:sect}

We define
\begin{equation}
\mathbf{A}=\left[ \int_0^{\infty} \mathbf{C}(\tau) d\tau \right]^{\text{T}}
\label{Adefn:eqn}
\end{equation}
and rearrange (\ref{gaussianFDT:eqn}) to get
\begin{equation}
\mathbf{A} \mathbf{B}^{\text{T}} = -\mathbf{C}(0).
\label{invereSystemFull:eqn}
\end{equation}
A row vector $\mathbf{b}_i$ may be defined so that it contains the $i$th row and only the $r$ non-zero columns of $\mathbf{B}$. $r$ is equal to the number of elements of $\mathbf{B}$ that exist in the expression for the right hand side of $dx_i/dt$, where $x_i$ denotes a single element of the vector $\mathbf{x}$. $\mathbf{A}$ may be truncated to a matrix $\mathbf{A}'_i$ that contains all $d$ rows and $r$ columns of $\mathbf{A}$. The index of each included column corresponds to the non-zero columns of $\mathbf{B}$ in its $i$th row.
% See figure \ref{smallerB:fig}.
Then for each column $i=1...d$ of $\mathbf{C}(0)$, denoted $\mathbf{c}_i$, we may write
\begin{equation}
\mathbf{A}'_i \mathbf{b}_i^{\text{T}}=-\mathbf{c}_i.
\label{invereSystemSingle:eqn}
\end{equation}
For a finite number of data points, $n$, any estimate of $\mathbf{C}(\tau)$ or $\mathbf{A}$ using data, (\ref{CtauFromData:eqn}), will typically result in an overestimate of the magnitude of elements where the true value is sufficiently close to zero. Contribution of this overestimate to any resulting estimate of $\mathbf{b}_i$ will in general depend upon $d$. Assuming that $\mathbf{C}(\tau)$ decays with distance, we pick the distances $a_{\text{cut}}$ and $c_{\text{cut}}$ above which the respective elements of $\mathbf{A}$ and $\mathbf{C}(0)$ are set to zero and call the truncated version of this matrix and vector $\mathbf{A}''_i$ and $\mathbf{c}''_i$. This gives
\begin{equation}
\mathbf{A}''_i \mathbf{b}_i^{\text{T}} \approx -\mathbf{c}''_i.
\label{localGaussianFDT:eqn}
\end{equation}
All of the $\mathbf{b}_i$'s may then be found
% using the psudeo inverse (\cite{MatrixCookbook})
%
%\begin{equation}
%\mathbf{B}'_i \approx \mathbf{A}''_i^+ \mathbf{c}''_i,
%\end{equation}
%
%i.e.
using a linear least squares fit via the singular value decomposition, and combined to give an estimate of $\mathbf{B}$. We call this method the local Gaussian FDT because we have made the approximation that perturbations decay to zero after some distance. The local Gaussian FDT may be generalised to the non-linear, non-Gaussian case in the context of estimating the response to a forcing \cite{Cooper11}.

\subsection{Local LIM}
\label{localLinearFDT:sect}

If we instead make the approximation that information travels at a finite speed between different $x_i$, then after a sufficiently short time $\tau$ the set of elements $\mathbf{x}'''_i$ of $\mathbf{x}$ that can possibly have an important influence upon $x_i$ are those corresponding to the non-zero elements of row $i$ of $\mathbf{B}$. As $\tau$ increases, the number of elements of $\mathbf{x}$ that have an important influence upon $x_i$ increases. We therefore assume that for small $\tau$, (\ref{linearFDT:eqn}) does not require information from the full covariance matrices $\mathbf{C}(0)$ and $\mathbf{C}(\tau)$ in order to accurately approximate $\mathbf{B}$. The accuracy of this approximation for any given $\tau$ depends upon $\mathbf{B}$. To estimate row $i$ of $\mathbf{B}$ the covariance matrices $\mathbf{C}'''_i(0)$ and $\mathbf{C}'''_i(\tau)$ of the time series $\mathbf{x}'''_i$ are required
\begin{equation}
\mathbf{B}'''_i=\frac{1}{\tau}\log\left[ \mathbf{C}'''_i(\tau) \mathbf{C}'''_i(0)^{-1} \right].
\label{localLinearFDT:eqn}
\end{equation}
Then the $\mathbf{b}_i$, (defined in section \ref{localGaussianFDT:sect}), used to estimate $\mathbf{B}$, are given by the corresponding row of $\mathbf{B}'''_i$ for each $i$. We call this method local linear inverse modelling because we have made the approximation that perturbations can only be felt within a finite distance after a finite time.

\subsection{The noise covariance matrix.}

The time taken for a $d\times d$ dimensional matrix matrix multiplication is conventionally proportional to $d^3$. However if $\mathbf{B}$ is sparse
%and since $\mathbf{C}(0)$ is symmetric with non-zero off diagonal elements,
the time taken to find $\mathbf{Q}$ using (\ref{Lyapunov:eqn}) can be faster, proportional to $d^2$. $\mathbf{Q}$ is in general dense, even if $\mathbf{B}$ is sparse, so finding $\mathbf{Q}$ may be problematic for extremely large $d$. It turns out that finding all elements of $\mathbf{Q}$ may not be necessary. If we assume that the process generating the noise $\bm{\xi}$ is somewhat local in nature, then both the number of elements of $\mathbf{Q}$ necessary for numerical integration of (\ref{linearModel:eqn}) and the time taken to generate $\bm{\xi}(t)$ at a particular time $t$ is proportional to $d$, see the appendix.

%%%%%%%%%
% New Section %
%%%%%%%%%

\section{Test integrations}
\label{testModel:sect}

We compare algorithms by testing them with a time series generated by a simple linear stochastic model. The model we use is a system of coupled linear equations
\begin{equation}
\frac{\text{d}x_i}{\text{d} t} = \left( b_{i,i-1} x_{i-1} + b_{i,i} x_i + b_{i,i+1} x_{i+1} \right) +\xi_i\label{testSystem:eqn}
\end{equation}
with $i=1 \dots d$. $b_{i,j}$ represents the elements of the constant matrix $\mathbf{B}$ with $b_{1,0}=b_{d,d+1}=0$ and all elements of $\mathbf{B}$ not included in (\ref{testSystem:eqn}) being zero. The deterministic part of this system is similar in nature to a discretised partial differential equation with one spatial and one temporal dimension. The constants $b_{i,j}$ are chosen randomly using the following expressions
%
% As in the C program:
%\begin{align*}
%b_{i,i} &= -\left( 1-\alpha \right) u_{i,1} - \alpha, & 1 \leq i \leq d, \\
%b_{i,i+1} &= -\left( u_{i,2} \left( 1-\beta \right) + \beta \right) b_{i,i}, & %1 \leq i \leq d-1, \\
%b_{i,i-1} &= -\left( 1-\alpha \right) u_{i,3} \left( b_{i,i}+b_{i,i+1} \right), %& 2 \leq i \leq d,
%\end{align*}
%
% Simplified:
\begin{align*}
b_{i,i} &= \left( \alpha-1 \right) u_{i,1} - \alpha, & 1 \leq i \leq d, \\
b_{i,i+1} &= \left( \left( \beta -1 \right) u_{i,2} - \beta \right) b_{i,i}, & 1 \leq i \leq d-1, \\
b_{i,i-1} &= \left( \alpha-1 \right) u_{i,3} \left( b_{i,i}+b_{i,i+1} \right), & 2 \leq i \leq d,
\end{align*}
where $u_{i,j}$ represents a random number chosen from a uniform distribution between 0 and 1, $\alpha=0.5$ and $\beta=0.35$ are constants that govern the autocorrelation decay time and the coupling between state vector elements respectively. In our test case we assume that only the tridiagonal elements of $\mathbf{Q}$ are important. Then since $\mathbf{Q}$ is symmetric, only the diagonal and +1 off diagonal elements are required. For the hypothetical physical system that we are assuming, the other elements of $\mathbf{Q}$ add no additional useful information. The diagonal of $\mathbf{Q}$ is then chosen by
\[
q_{i,i}=\frac{1}{3} \left( r_{i,1}^2 + r_{i,2}^2 + r_{i,3}^2 \right), \qquad 1 \leq i \leq d,
\]
and the +1 off diagonal is given by
\[
q_{i,i+1}=\frac{1}{3} \left( r_{i,1} r_{i+1,1} + r_{i,2} r_{i+1,2} + r_{i,3} r_{i+1,3} \right), \qquad 1 \leq i \leq d-1,
\]
where $r_{i,j}$ represents a random number chosen from a Gaussian distribution with zero mean and unit variance. (\ref{testSystem:eqn}) is integrated for each $i$ using the Euler-Maruyama method
\[
x_i^{n+1} = x_i^n + \left( b_{i,i-1} x_{i-1}^n + b_{i,i} x_i^n + b_{i,i+1} x_{i+1}^n \right) \Delta t + \xi_i^n \sqrt{\Delta t}, \qquad 1 \leq i \leq d.
\]
where $\Delta t=0.01$ is chosen as the time step, the state vector is recorded at each time step, and the values of $\xi_i^n$ are chosen given the method in the appendix. This test system is particularly simple, so in this case
\[
\xi_1^n=\sqrt{q_{1,1}} \psi_1^n
\]
and
\[
\xi_i^n = \frac{q_{i-1,i}}{q_{i-1,i-1}} \xi_{i-1}^n + \sqrt{ q_{i,i}-\frac{q_{i-1,i}^2}{q_{i-1,i-1}} } \psi_i^n, \qquad 2 \leq i \leq d
\]
where $\psi_i^n$ are random numbers chosen from a Gaussian distribution with zero mean and unit variance.

When applying the Gaussian FDT (\ref{gaussianFDT:eqn}) and local Gaussian FDT, (\ref{Adefn:eqn}) and (\ref{localGaussianFDT:eqn}), to any of these data sets, an upper limit of the integral of 20 is chosen to approximate infinity. When applying LIM (\ref{linearFDT:eqn}), or local LIM (\ref{localLinearFDT:eqn}) to any of these data sets, a value of $\tau=1$ is chosen. $n$ instances of the state vector at lag zero and lag one and the integral of the state vector are kept every 20 time units. All intermediate data and a spin up from $t=-20$ to $t=0$ with random initial conditions, is discarded.

%%%%%%%%%
% New Section %
%%%%%%%%%

\section{Results}
\label{results:sect}

\subsection{Clipping the local Gaussian FDT}

\begin{figure}
\begin{center}
\includegraphics[width=0.49\textwidth]{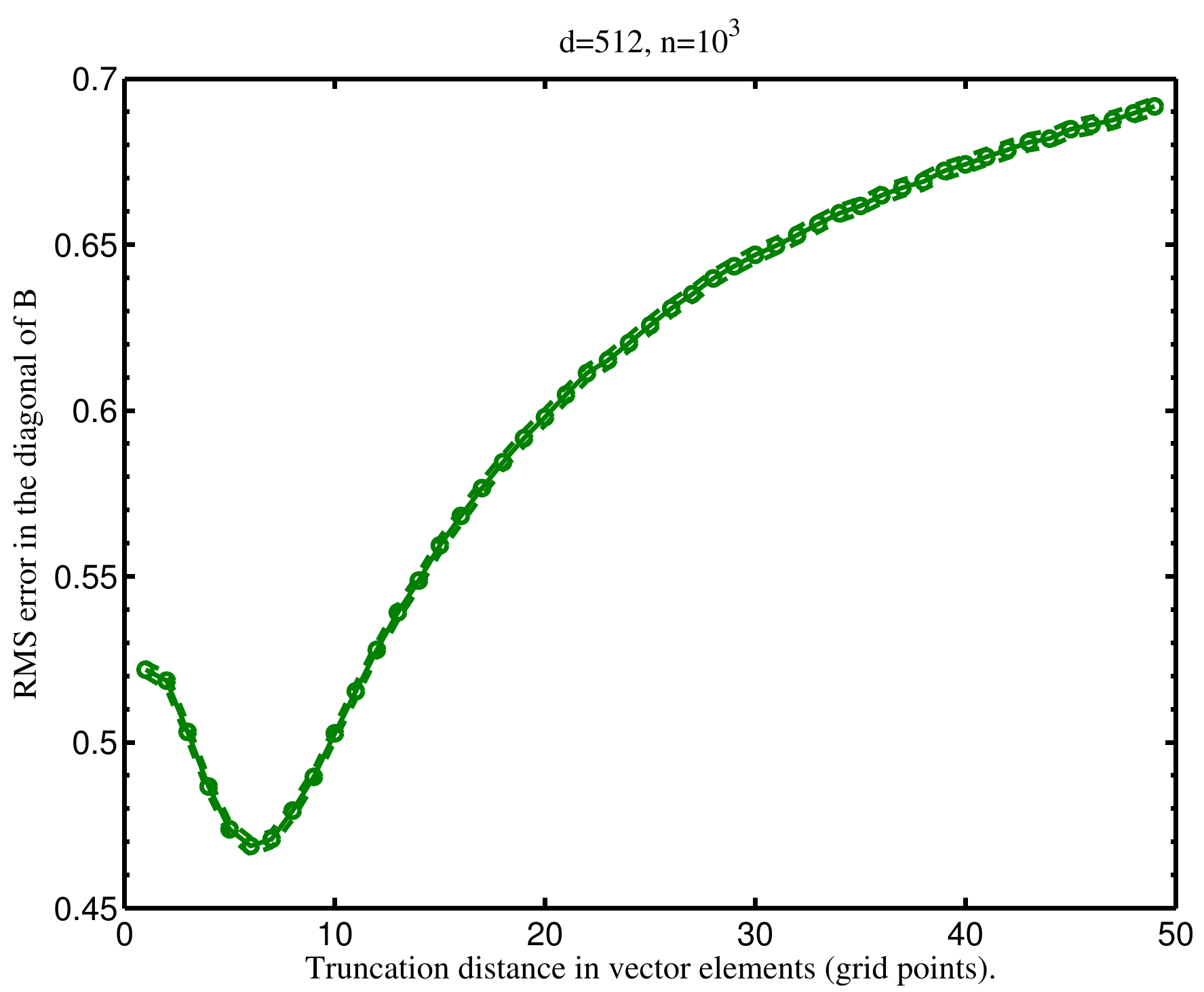}
\includegraphics[width=0.49\textwidth]{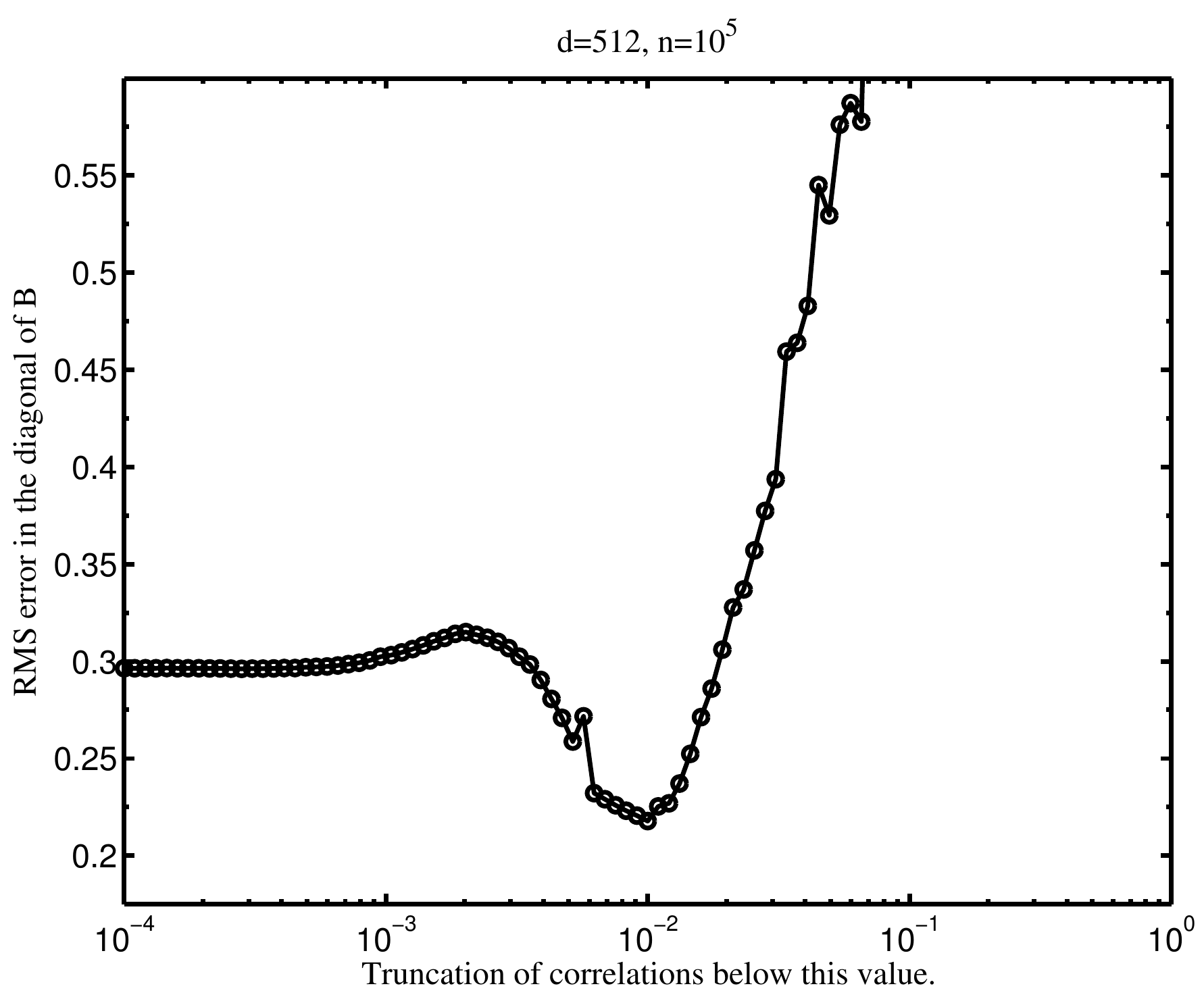}
\end{center}
\caption{A plot of the RMS error in the diagonal of $\mathbf{B}$ as a function of the distance (left) and magnitude (right) of ignored correlations for the local Gaussian FDT with $d=512$. For truncation with distance (left), the circles represent the ensemble mean over 100 independent data sets each of $n=1000$ data points. The dashed lines denote the ensemble standard deviation multiplied by $2/\sqrt{100}$. For truncation by correlation, a single data set of $n=10^5$ instances of the state vector was used. }
\label{corrClip:fig}
\end{figure}

For the local Gaussian FDT we need to truncate the matrix $\mathbf{A}$ from (\ref{Adefn:eqn}) and the covariance matrix $\mathbf{C}(0)$. If our physical understanding of the system is not sufficient then, since we have a noisy estimate of these matrices from the data, this cut-off can be chosen by looking at their structure and an add-hoc estimate of a zero correlation distance. To illustrate the gains in accurate estimation of $\mathbf{B}$ by assuming a cut-off, the root mean squared (RMS) error is plotted as a function of the clipping distance in figure \ref{corrClip:fig}. An alternative method, also plotted in figure \ref{corrClip:fig}, is to choose to cut-off at a particular correlation. We choose a minimum allowed correlation and set elements of $\mathbf{A}$ and $\mathbf{C}(0)$ to zero, where the correlation is below its cut-off value. Figure \ref{corrClip:fig} shows that the local Gaussian FDT performs a more accurate estimate for a moderate cut-off. Ignoring distances above 6 vector elements or correlations below 0.01 is close to optimal in this case. Some form of cross validation can be used to choose the cut-off when applying the local Gaussian FDT to a new data set. Similar results are obtained for the RMS error in the $\pm 1$ off diagonals. The optimum truncation depends upon several factors and for simplicity we choose to truncate all lag-correlations (to zero) above a distance of 32 vector elements in all further applications of the local Gaussian FDT in this paper.

\subsection{Convergence with length of the data set}

\begin{figure}
\begin{center}
\includegraphics[width=0.49\textwidth]{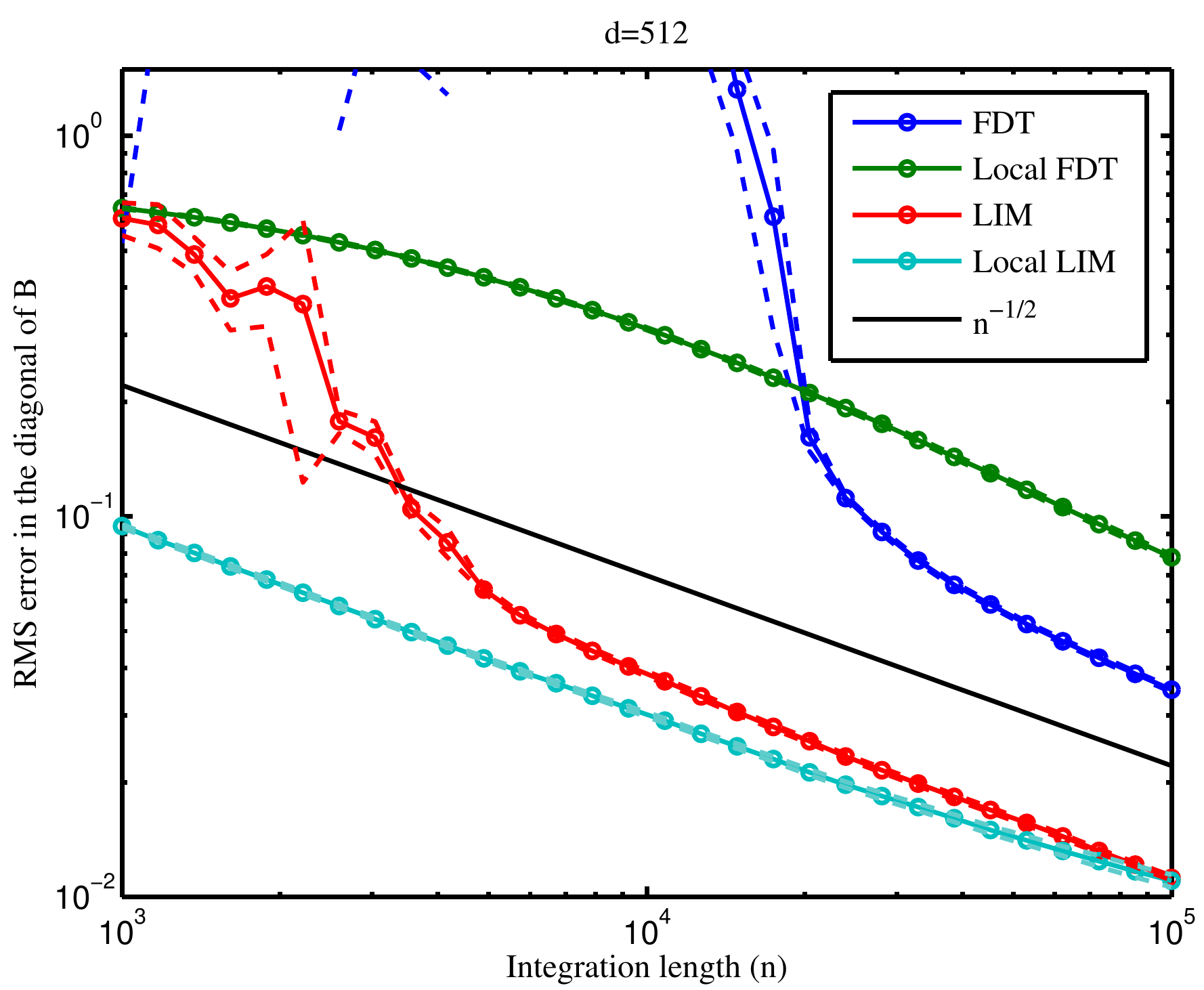}
\end{center}
\caption{Accuracy of each method as a function of the time window over which data is available. Note the logarithmic axes. For these integrations $d=512$. The circles and solid lines indicate the mean over an ensemble of 100 independent members. The dashed lines indicate the ensemble standard deviation multiplied by $2/\sqrt{100}$.}
\label{RMSvRunLength:fig}
\end{figure}

Correlations of a stable linear stochastic system decay exponentially in time. So after some time the system is effectively independent of its initial state. Therefore the expected error in the mean calculated from a sample of size $n$ of independent random numbers is proportional to $1/\sqrt{n}$. For sufficient data, we expect the error in estimates of $\mathbf{B}$ and $\mathbf{Q}$ to also be proportional to $1/\sqrt{n}$. This is examined in figure \ref{RMSvRunLength:fig} which shows that given sufficient data, the error in all methods appears to decay approximately according to this rule. The error in the FDT demonstrates the quantity of data required before reasonable estimates can be obtained (around 2 to $3 \times 10^4$ data points for $d=512$). In comparison, the local FDT has good performance for small data sets but is beaten by the FDT for large data sets. This reflects the fact that correlations beyond the 32 grid point clipping distance, are resolved. In this test LIM without truncation performs better than both forms of the FDT. Like the FDT, a certain quantity of data is required to dramatically reduce the error before $1/\sqrt{n}$ behaviour is approximately reached. The local LIM method yields the smallest error, at large $n$ having effectively the same accuracy as the LIM method without truncation. Similar results were obtained for the off diagonal elements of $\mathbf{B}$ and the elements of $\mathbf{Q}$.

\subsection{Accuracy as a function of dimensionality}

\begin{figure}
\begin{center}
\includegraphics[width=0.49\textwidth]{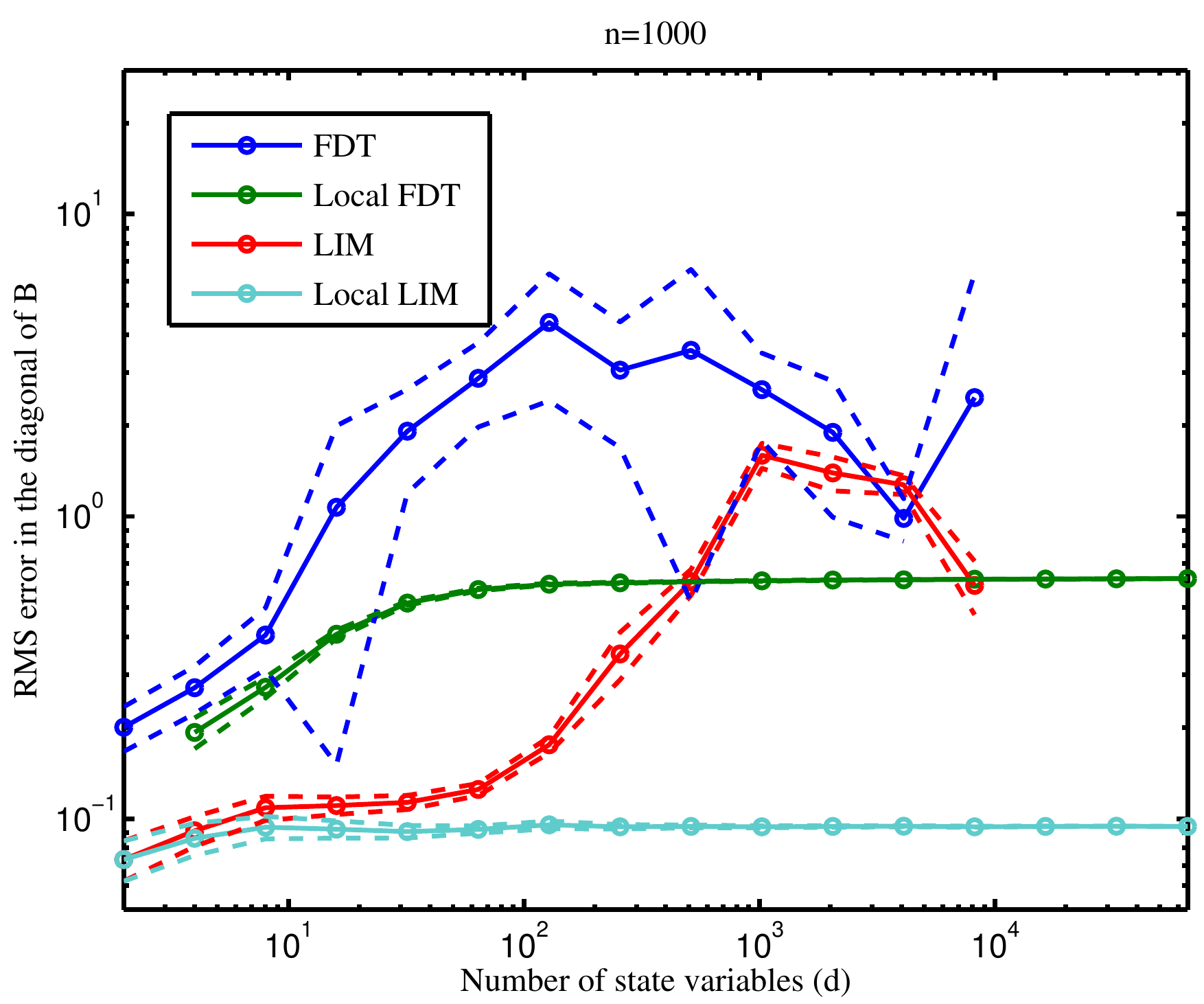}
\end{center}
\caption{The accuracy of each method as a function of dimensionality of the state vector. Note the logarithmic axes. Each estimate uses $n=1000$ instances of the state vector with the highest dimensionality tested being $d=65536$ ($=2^{16}$). The circles and solid lines indicate the mean over an ensemble of 100 independent members. The dashed lines indicate the ensemble standard deviation multiplied by $2/\sqrt{100}$. The highest dimensional points for the conventional FDT and LIM are omitted due to computational time and memory limitations.}
\label{RMSvsD:fig}
\end{figure}

For a limited data set of 1000 points in time, figure \ref{RMSvsD:fig} demonstrates the error of each method as a function of dimensionality $d$ of the state vector $\mathbf{x}$. The FDT and local FDT display similar performance for low dimensional ($d<10$) systems, but the performance of the FDT becomes poor and unpredictable at higher dimensionality while the performance of the local FDT plateaus. Similar behaviour is observed for LIM and local LIM. It is clear that for $d>100$ the performance of the local FDT and local LIM is independent of $d$. Again, similar results were obtained for the off diagonal elements of $\mathbf{B}$ and the elements of $\mathbf{Q}$.

\subsection{Computation time as a function of dimensionality}

\begin{figure}
\begin{center}
\includegraphics[width=0.49\textwidth]{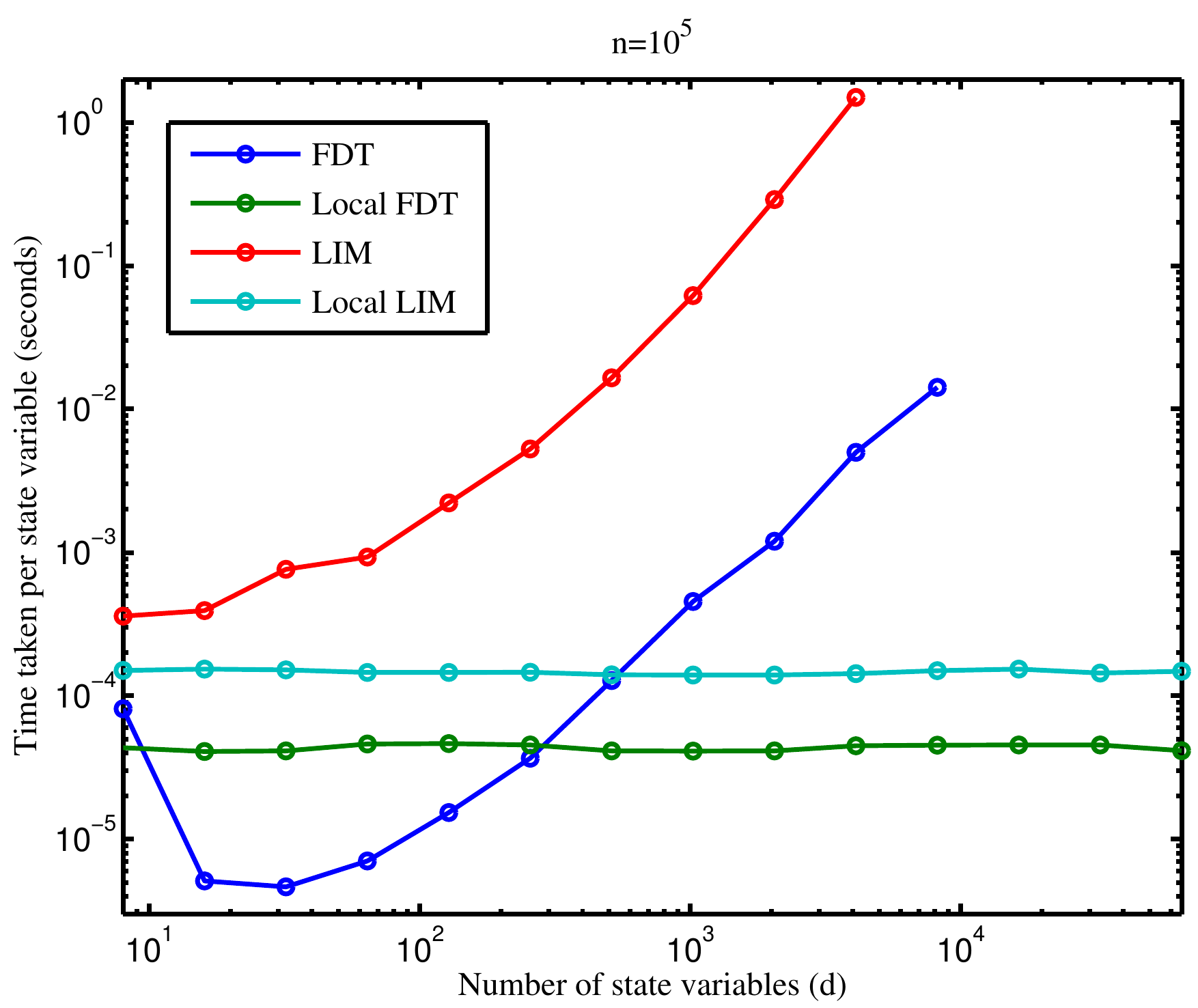}
\end{center}
\caption{The time taken for each method per state vector element as a function of dimensionality of the state vector. Note the logarithmic axes. Calculations and timings were performed on a standard desktop computer and each estimate was performed with a single integration providing $n=10^5$ instances of the state vector.}
\label{timeVsD:fig}
\end{figure}

Figure \ref{timeVsD:fig} shows the time taken per variable for the application of each method as a function of dimensionality. The time taken for the Gaussian FDT is dominated by the matrix inverse operation and the time taken for LIM is dominated by the matrix logarithm operation. The time taken for either is proportional to $d^3$ using conventional algorithms. The time taken for their local counterparts is proportional to $d$ in both cases.

%%%%%%%%%
% New Section %
%%%%%%%%%

\section{Conclusion}
\label{conclusions:sect}

We have presented two algorithms for finding the parameters governing a high dimensional linear system. We call them the local Gaussian fluctuation dissipation theorem (FDT) and local linear inverse modelling (LIM). The accuracy of these algorithms does not depend upon the dimensionality of the state vector and the time taken in practice for their computation is proportional to the number of state vector elements. We have tested our algorithm with linear stochastic systems of up to $2^{16}$ variables. Although the particular application in mind here is approximation of a turbulent fluid, we expect that this method can be applied in other contexts.

The conventional method of dealing with high dimensional systems is to first make a truncation into some smaller space, for example EOF space. Unfortunately it may be the case that cut-off in the spatial spectrum of a turbulent fluid leads to less accurate predictions. The algorithms presented do not require such a truncation, even for extremely high dimensional systems. Instead, some kind of locality of the system needs to be assumed. For example, in a spatial discretisation of a field into a number of grid points, that reasonable perturbations to a single grid point only affect a limited number of local grid points after a small amount of time. After longer times, the fact that a perturbation can propagate over the entire domain does not reduce the accuracy of the local linear FDT. However, the accuracy and computation time of the local Gaussian FDT depends upon the number of grid points that the perturbation propagates over. As is standard with the discretisation of many partial differential equations, part of this assumption requires that the matrix $\mathbf{B}$ in equation (\ref{linearModel:eqn}) is sparse.

In addition to the choice of a lag time that approximates infinity, required by the standard Gaussian FDT, the local Gaussian FDT requires the specification of the maximum distance that a perturbation can propagate before becoming insignificant. This quantity may typically be estimated by understanding of the physical system modelled and by looking at the spatial range of typical correlations in the data. For the local linear FDT, the state vector elements that a perturbation can reach after a sufficiently small time, for example one model time step, must also be provided. Both the standard linear FDT and the local linear FDT require the choice of a parameter that represents this sufficiently small time.

%In addition we have presented in the appendix a fast method of generating fields of spatially correlated random numbers. The spatial correlation is local and is easily chosen directly by the user to be any reasonable spatially varying value. The motivation here was to enable fast integration of large linear stochastic systems, however the method is also applicable elsewhere.

\appendix


%%%%%%%%%
% New Section %
%%%%%%%%%

\section{Generation of spatially correlated random numbers.}
\label{RandomTerm:sect}

\begin{figure}
\begin{center}
\includegraphics[width=0.2\textwidth]{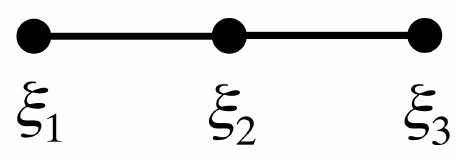}
\end{center}
\caption{An example of what we mean by a direct link between the physical process that underly the white noise process. $\xi_1$ is directly linked by some physical process to $\xi_2$. $\xi_2$ is directly linked to both $\xi_1$ and $\xi_3$. $\xi_3$ is directly linked to $\xi_2$. Although $\xi_1$ and $\xi_3$ may be correlated, it is only due to their shared correlation with $\xi_2$. We assume that there is no additional process that links them providing additional correlation. They are therefore not directly linked and the correlation between them provides no additional useful information.}
\label{directLink:fig}
\end{figure}

Numerical integration of a single realisation of (\ref{linearModel:eqn}) requires generation of correlated random numbers representing $\bm{\xi}$. To do this in the conventional way, $\mathbf{Q}$ must be found using (\ref{Lyapunov:eqn}). The noise at time $t$ may then be generated by
\begin{equation}
\bm{\xi}(t)=\mathbf{V} \sqrt{\mathbf{D}} \bm{\psi}(t)
\label{conventionalNoise:eqn}
\end{equation}
where $\mathbf{V}$ is the matrix of the eigenvectors of $\mathbf{Q}$, $\mathbf{D}$ is the corresponding diagonal matrix of eigenvalues, the square root is taken element wise and $\bm{\psi}$ is a vector of Gaussian distributed independent random numbers with unit variance. However the computational time required to estimate $\mathbf{V}$ and $\mathbf{D}$ is proportional to $d^3$ and the computational time and memory required to evaluate (\ref{conventionalNoise:eqn}) is proportional to $d^2$ unless $\mathbf{V}$ has a simple structure. By making the assumption that element $j$ of $\bm{\xi}$, $\xi_j$, is only directly linked to $p_j$ other elements, the computational time and memory required may be reduced to be proportional to $\sum_{j=1}^d p_j^3$ and $\sum_{j=1}^d p_j^2$ respectively. It is possible that correlations between $\xi_1$ and $\xi_3$ are entirely caused by both of them being correlated with $\xi_2$. So in this context ``direct link'' means that correlations between $\xi_1$ and $\xi_3$ do not require another $\xi_i$, see figure \ref{directLink:fig}. Which of the elements of $\bm{\xi}$ are directly linked to each other is a necessary assumption and may be justified, for example, by understanding of the physical process modelled.

\subsection{A faster method}

The first element of $\bm{\xi}$, $\xi_1$, may be generated at any time by
\begin{equation}
\xi_1(t) = \sqrt{a_{1,1}} \psi_1(t)
\label{firstNoisePt:eqn}
\end{equation}
where $\psi_j(t)$ is an element of $\bm{\psi}(t)$, a Gaussian distributed random number with unit variance and zero temporal correlation. The constant $a_{1,1}$ is the variance of $\xi_1$ given by $a_{1,1}=q_{1,1}$ with $q_{i,j}$ representing element $i,j$ of $\mathbf{Q}$. If correlations in the next point, $\xi_2$, with $\xi_1$ are caused by a direct link between the physical processes governing these two terms then
\[
\xi_2(t) = a_{1,2} \xi_1(t) + \sqrt{a_{2,2}} \psi_2(t).
\label{xi2:eqn}
\]
Here $\xi_2$ has a component correlated with $\xi_1$ with a magnitude given by the constant $a_{1,2}$ and a component that is independent of $\xi_1$ with a magnitude given by $a_{2,2}$. $a_{1,2}$ must be chosen so that the covariance of $\xi_1$ and $\xi_2$ equals $q_{1,2}$. Therefore
\begin{align*}
\left< \xi_1 \xi_2 \right> &= q_{1,2} \\
\left< \xi_1 \left( a_{1,2} \xi_1 + \sqrt{a_{2,2}} \psi_2 \right) \right> &= q_{1,2} \\
a_{1,2} \left< \xi_1 \xi_1 \right> &= q_{1,2} \\
a_{1,2} q_{1,1} &= q_{1,2}.
\end{align*}
$a_{2,2}$ must now be chosen so that the variance of $\xi_2$ is equal to $q_{2,2}$,
\begin{align*}
\left< \xi_2 \xi_2 \right> &= q_{2,2} \\
\left< \left( a_{1,2} \xi_1 + \sqrt{a_{2,2}} \psi_2 \right) \left( a_{1,2} \xi_1 + \sqrt{a_{2,2}} \psi_2 \right) \right> &= q_{2,2} \\
\left< a_{1,2}^2 \xi_1^2 + 2 a_{1,2} \sqrt{a_{2,2}} \xi_1 \psi_2 + a_{2,2} \psi_2^2 \right> &= q_{2,2} \\
a_{1,2}^2 \left< \xi_1^2 \right> + a_{2,2} \left< \psi_2^2 \right> &= q_{2,2} \\
a_{1,2}^2 q_{1,1} + a_{2,2} &= q_{2,2}.
\end{align*}
Therefore
\[
a_{2,2}=q_{2,2}-a_{1,2}^2 q_{1,1}.
\]
Written as a matrix equation and noting that $q_{i,j}=q_{j,i}$, $a_{1,1}$ and $a_{2,2}$ may be found by solving
\[
\left( \begin{array}{cc}
q_{1,1} & 0 \\
q_{1,2} & 1
\end{array} \right)
\left( \begin{array}{c}
a_{1,2} \\
a_{2,2}
\end{array} \right) =
\left( \begin{array}{c}
q_{1,2} \\
q_{2,2}
\end{array} \right).
\]
If the element $\xi_3$ is directly linked with $\xi_1$ and $\xi_2$ then
\[
\xi_3(t) = a_{1,3} \xi_1(t) + a_{2,3} \xi_2(t) + \sqrt{a_{3,3}} \psi_3(t)
\]
and the same reasoning as before leads to the matrix equation
\[
\left( \begin{array}{ccc}
q_{1,1} & q_{2,1} & 0 \\
q_{1,2} & q_{2,2} & 0 \\
q_{1,3} & q_{2,3} & 1
\end{array} \right)
\left( \begin{array}{c}
a_{1,3} \\
a_{2,3} \\
a_{3,3}
\end{array} \right) =
\left( \begin{array}{c}
q_{1,3} \\
q_{2,3} \\
q_{3,3} \\
\end{array} \right)
\]
which must be solved for $a_{1,3}$, $a_{2,3}$ and $a_{3,3}$. Similarly if $\xi_4$ is directly linked with  $\xi_1$, $\xi_2$ and $\xi_3$ then
\[
\xi_4(t) = a_{1,4} \xi_1(t) + a_{2,4} \xi_2(t) + a_{3,4} \xi_3(t) + \sqrt{a_{4,4}} \psi_4(t)
\]
and we get
\[
\left( \begin{array}{cccc}
q_{1,1} & q_{2,1} & q_{3,1} & 0 \\
q_{1,2} & q_{2,2} & q_{3,2} & 0 \\
q_{1,3} & q_{2,3} & q_{3,3} & 0 \\
q_{1,4} & q_{2,4} & q_{3,4} & 1 \\
\end{array} \right)
\left( \begin{array}{c}
a_{1,4} \\
a_{2,4} \\
a_{3,4} \\
a_{4,4} \\
\end{array} \right) =
\left( \begin{array}{c}
q_{1,4} \\
q_{2,4} \\
q_{3,4} \\
q_{4,4} \\
\end{array} \right).
\]
On the other hand, if $\xi_4(t)$ is not directly related to $\xi_1(t)$, only being directly related to $\xi_2(t)$ and $\xi_3(t)$, then
\begin{equation}
\xi_4(t) =a_{2,4} \xi_2(t) + a_{3,4} \xi_3(t) + \sqrt{a_{4,4}} \psi_4(t)
\label{indepNoise:eqn}
\end{equation}
and
\[
\left( \begin{array}{ccc}
q_{2,2} & q_{3,2} & 0 \\
q_{2,3} & q_{3,3} & 0 \\
q_{2,4} & q_{3,4} & 1
\end{array} \right)
\left( \begin{array}{c}
a_{2,4} \\
a_{3,4} \\
a_{4,4}
\end{array} \right) =
\left( \begin{array}{c}
q_{2,4} \\
q_{3,4} \\
q_{4,4} \\
\end{array} \right).
\]
The precise numerical values of the indices may be altered to suit a particular problem without difficulty. Although we have only made use of a subset of the elements of $\mathbf{Q}$, it is not sparse. If necessary, any of the remaining elements of $\mathbf{Q}$ may be found in terms of elements already found. For example $q_{1,4}$ using equation (\ref{indepNoise:eqn}) is given by
\begin{align*}
 q_{1,4} &= \left< \xi_1 \xi_4 \right> \\
&= \left< \xi_1 \left( a_{2,4} \xi_2 + a_{3,4} \xi_3 + \sqrt{a_{4,4}}\psi_4 \right) \right> \\
&=a_{2,4} \left< \xi_1 \xi_2 \right> + a_{3,4} \left< \xi_1 \xi_3 \right> \\
&=a_{2,4} q_{1,2} + a_{3,4} q_{1,3}.
\end{align*}
Note that these values, calculated for remaining undefined elements of $\mathbf{Q}$, are only valid for the specific order of the equations used to generate $\bm{\xi}$. If a different element of $\bm{\xi}$ is used as a starting point (\ref{firstNoisePt:eqn}), then in general different values of the undefined elements of $\mathbf{Q}$ may be found. We are assuming that either the differences are not important or that that the starting point is somehow correctly chosen.

\section*{Acknowledgements}
This work was funded by UK NERC grant NE/J00586X/1.


\bibliographystyle{elsarticle-num} 
\bibliography{Cooper2014.bbl}


%\begin{thebibliography}{00}


%\bibitem{}

%\end{thebibliography}
\end{document}